\documentclass{phyeauth}

\usepackage{graphicx}
\usepackage{amsmath}
\usepackage{amssymb}
\usepackage{bm}% bold math
\usepackage{bbm}
\usepackage{psfrag}

\newcommand{\I}{\textup{i}}

\newcommand{\pop}[2]{\frac{\partial #1}{\partial #2}}

\newcommand{\ddim}{\udelta\kern0.1em}

\newcommand{\beikonst}[2]{\left( #1 \right)_{\kern-0.2em #2}}
\newcommand{\tr}[2][]{\text{Tr}_{#1}\left\{#2\right\}}
\newcommand{\trtxt}[2][]{\text{Tr}_{#1}\{#2\}}
\newcommand*{\ketL}[1]{\mathopen{|}#1\mathclose{)}}
\newcommand*{\braL}[1]{\mathopen{(}#1\mathclose{|}}
\newcommand*{\sprodL}[2]{\mathopen{(}#1 |#2 \mathclose{)}}

\begin{document} 

% -----------------------------------------
%
% Title
%
\begin{frontmatter}

\title{Quantum Heat Transport: Perturbation Theory in Liouville Space} 

\author[stg]{Mathias Michel\corauthref{cor}},
\corauth[cor]{Corresponding author.}
\ead{mathias@theo1.physik.uni-stuttgart.de}
\author[osn]{Jochen Gemmer} and
\author[stg]{G\"unter Mahler}

\address[stg]{Institut f\"ur Theoretische Physik I, Universit\"at Stuttgart, 70550 Stuttgart, Germany}
\address[osn]{Fachbereich Physik, Universit\"at Osnabr\"uck, 49069 Osnabr\"uck, Germany}

\date{\today}% 
\begin{abstract}
We consider chains consisting of several identical subsystems weakly coupled by various types of next neighbor interactions.
At both ends the chain is coupled to a respective heat bath with different temperature modeled by a Lindblad formalism.
The temperature gradient introduced by this environment is then treated as an external perturbation.
We propose a method to calculate the heat current and the local temperature profile of the resulting stationary state as well as the heat conductivity in such systems.
This method is similar to Kubo techniques used e.g.\ for electrical transport but extended here to the Liouville space.
\end{abstract}

\end{frontmatter}

% ---------------------------------------------------------------------------
%
% Main text
%
% ---------------------------------------------------------------------------

\section{Introduction}
\label{sec:level1}

As a specific topic of non-equilibrium thermodynamics, heat conduction has long since been of central interest.
Instead of reaching a complete equilibrium state, the composite system under some appropriate perturbation enters a local equilibrium state -- small parts of the system approach equilibrium but not the whole system.

Within non-equilibrium statistical mechanics the theory of linear reponse, originally developed to account for electric conductivity, is a very import method to investigate dynamical as well as static properties of materials \cite{Kubo1991,Mori1956,Mahan1981,Zubarev1974}.
In this context the famous Kubo-formulas \cite{Kubo1957} have led to a rapid developement in the theoretical understanding of processes induced by an external perturbation of the system.
However, a direct mapping of these ideas on pure thermal transport phenomena (perturbations due to thermal gradients \cite{Luttinger1964}) faces serious problems:
Contrary to the case of external perturbations by an electric field, thermal perturbations cannot directly be described by a potential term in the hamiltonian of the system.
Rather, the thermal perturbation is introduced by heat baths with different temperatures coupled to the system, thus calling for a more detailed description than is needed for electric transport.
Nevertheless, those methods are often used, eventually because of their immediate success in describing non-equilibrium processes \cite{Kubo1991,Zotos1997,Heidrich2003,Kluemper2002}.

Recently, the main focus of considerations on heat conduction and Fourier's law has shifted towards small (one dimensional) quantum systems \cite{Saito2003,Saito1996}.
Typically, these systems are chains of identical subsystems weakly coupled by some next neighbor interaction. 
Based on the Lindblad formalism \cite{Lindblad1976} or on techniques of quantum master equations \cite{Saito2002}, heat baths are then weakly coupled to the chain at both ends.
It has been found that in such systems the appearance of a normal heat conduction depends on the type of the interaction between these elementary subsystems \cite{Michel2003}.
Most quantum mechanical interaction types show a normal heat conduction behavior (constant non-vanishing local temperature gradients), whereas for some special coupling the local gradient within the chain vanishes (divergence of the conductivity, non-normal scenario).

Let us now introduce the model system, which we are going to investigate in the following based on the full numerical integration of Liouville-von Neumann equation as well as on a perturbation theory in Liouville Space.

% ---------------------------------------------
%
% Second Chapter:
%
\section{Model System}
\label{sec:level2}

The dynamics of the quantum model for heat transport is given by the Liouville-von-Neumann equation (LvN) for open systems
\begin{equation}
  \label{eq:8}
  \pop{}{t}\hat{\rho} = \mathcal{L}\hat{\rho}\;.
\end{equation}
Thus we have to consider super operators -- here $\mathcal{L}$ -- acting on operators in Hilbert space, e.g.\ the density operator of the system (see \cite{Schack2000,Tarasov2002,Mukamel2003}).   
The complete Liouville operator of the open system under consideration is given by
\begin{equation}
  \label{eq:1}
  \mathcal{L}= \mathcal{L}_{\text{sys}} 
                + \mathcal{L}_1(T_1) + \mathcal{L}_2(T_2)\;.
\end{equation}
The first term controls the coherent evolution of the quantum system defined by the Hamiltonian $\hat{H}$: 
It is defined by its action on the density operator $\hat{\rho}$ according to
\begin{equation}
  \label{eq:2}
  \mathcal{L}_{\text{sys}} \hat{\rho} 
  = -\frac{\I}{\hbar}\,[\hat{H},\hat{\rho}]\;,
\end{equation}
like for closed quantum systems.
The system $\hat{H}$ is here a chain of $N$ identical subunits with $n$ levels each, coupled weakly by a next neighbor interaction, thus living in a Liouville space of dimension $n^{2N}$.
One could think of several concrete model systems, for example spin models ($n=2$), for which the Hamiltonian would read
\begin{align}
  \label{eq:20}
  &\hat{H} = \sum_{\mu=1}^{N}\hat{\sigma}_3^{(\mu)}+
  \notag\\
  &\sum_{\mu=1}^{N-1}\big(
  J_x\hat{\sigma}_1^{(\mu)}\hat{\sigma}_1^{(\mu+1)}+
  J_y\hat{\sigma}_2^{(\mu)}\hat{\sigma}_2^{(\mu+1)}+
  J_z\hat{\sigma}_3^{(\mu)}\hat{\sigma}_3^{(\mu+1)}\big)\;.
\end{align}
The first term is the local part of the Hamiltonian, whereas the second defines the interaction between the subsystems ($\hat{\sigma}_i^{(\mu)}$ denote the Pauli operators of the $\mu$th spin).
Choosing $J_x = J_y = J_z$ we get the Heisenberg interaction and for $J_z=0$, $J_x=J_y$ an energy transfer coupling only (XY model).
Furthermore, to avoid any bias we will often use a random next neighbor interaction but without disorder (the same random interaction between different subsystems).

The chain is weakly coupled to two heat baths, one at each end of the system, given by the super operators $\mathcal{L}_1(T_1)$ and $\mathcal{L}_2(T_2)$ in (\ref{eq:1}).
This bath coupling could be realized by standard Lindblad operators \cite{Lindblad1976,Michel2003}, well known from the theory of open systems in quantum optics.
Another equivalent possibility is to derive a quantum master equation for the model system leading to a special bath coupling like e.g. in \cite{Saito2002}. 

% ---------------------------------------
%
% Third Chapter:
%
\section{Current and Local Temperature Profile}
\label{sec:level3}

We are interrested in the stationary state of the above described system, namely of the LvN equation (\ref{eq:8}) in heat conducting scenario. 
This stationary state - a local equilibrium state -- defines a temperture profile and a heat current.

As a measure for the temperature $T(\mu)$ of a single subsystem we use here the local energy of the respective system so that $0\leq T \leq 0.5$ in units of the local level spacing, c.f.\ \cite{Michel2003,Gemmer2004}.
This should be appropriate for weakly coupled subsystems within the chain (only a very small amount of energy is within the interaction).
The operator 
\begin{equation}
  \label{eq:31}
  \Delta\hat{H}^{(\mu,\mu+1)}_{\text{loc}} 
  = \hat{H}^{(\mu)}_{\text{loc}} - \hat{H}^{(\mu+1)}_{\text{loc}}\;,
\end{equation}
measures the local energy difference between two adjacent subsystems $\mu$ and $\mu+1$ ($\mu=1,2,\dots,N-1$).

The energy current operator $\hat{J}^{(\mu,\mu+1)}$ can be derived from a discretized version of the continuity equation (formulating a current into and out of the site $\mu$, respectively)
\begin{equation}
  \label{eq:32}
  \I [ \hat{H}, \hat{H}^{(\mu)}_{\text{loc}} ]
  = \hat{J}^{(\mu-1,\mu)} - \hat{J}^{(\mu,\mu+1)}\;.
\end{equation}

Both operators $\Delta\hat{H}^{(\mu,\mu+1)}_{\text{loc}}$, $\hat{J}^{(\mu,\mu+1)}$, will be used now to investigate the stationary state.

% ---------------------------------------------
%
% Forth Chapter:
%
\section{Perturbation Theory}
\label{sec:level4}

Beside a complete numerical investigation of the mentioned stationary state, let us try to get the local equilibrium state of the system out of a perturbation theory.
Since the system of the last section is living in Liouville space we have to work in this respective space, too.

Firstly, we consider the unperturbed system as our chain with both bath systems at the same temperature $T_1=T_2=T$.
Since the two baths have exactly the same temperature we expect the whole system to settle in a thermal stationary state $\hat{\rho}_0$:
This state should support neither a heat current nor temperature gradients -- it is a global equilibrium state with temperature $T$.

The eigenvalues and eigenvectors of the unperturbed system are given by the eigen equation
\begin{equation}
  \label{eq:4}
  \mathcal{L}_0 \ketL{\hat{\rho}_j} = l_j \ketL{\hat{\rho}_j}
  \;,\;j=0,\dots,n^{2N}-1\;,
\end{equation}
where the ket-vectors in Liouville space have been denoted as $\ketL{\dots}$.
A scalar product of vectors in Liouville space can be defined by
\begin{equation}
  \label{eq:25}
  \sprodL{\hat{\rho}_i}{\hat{\rho}_j}
  = \tr{\hat{\rho}_i^{\dagger}\hat{\rho}_j}\;.
\end{equation}
The (unique) stationary state $\hat{\rho}_0$ is also an eigenvector of the system with eigenvalue zero, $\mathcal{L}_0 \ketL{\hat{\rho}_0}=0$, whereas all other eigenvalues have a negative real part.
This is due to the fact that asymptotically the system should enter the unique equilibrium state $\ketL{\hat{\rho}_0}$ regardless of which state the system was at the beginning.
No other eigenvector is able to contribute to the equilibrium state, i.e.\ all other eigenvectors must be unstable.

Since the Liouville operator $\mathcal{L}_0$ is not hermitian, the eigenvectors do not form an orthogonal basis, i.e.\
\begin{equation}
  \label{eq:5}
  \sum_j \ketL{\hat{\rho}_j}\braL{\hat{\rho}_j}=\mathcal{G}
\end{equation}
is in general not the unit operator in Liouville space (c.f.\ \cite{Schack2000}).
But with the help of the super operator $\mathcal{G}$ it is possible to find a dual basis $\ketL{\hat{\rho}^j} = \mathcal{G}^{-1}\ketL{\hat{\rho}_j}$ with the property
\begin{equation}
  \label{eq:6}
  \sum_j \ketL{\hat{\rho}_j}\braL{\hat{\rho}^j} = \mathbbm{1}\;.
\end{equation}

The system will be perturbed now by applying a small temperature gradient $\Delta T$. 
We are interested in the properties of the stationary \emph{local} equilibrium state of the system reached because of this perturbation.
The Liouville operator of the perturbation is thus given by
\begin{equation}
  \label{eq:7}
  \mathcal{L'}(\Delta T) = \mathcal{L}_1(T+\frac{\Delta T}{2}) 
                            + \mathcal{L}_2(T-\frac{\Delta T}{2})\;.
\end{equation}
The two environment operators are the same as before but now with two different temperatures.

In the case of a chain of two level systems each of the two super operators $i=1,2$ of the bath coupling at both ends of the system consists of two transition processes (in case of finite temperatures)
\begin{equation}
  \label{eq:14}
  \mathcal{L}_i(T) = W_i^{\downarrow}(T) \mathcal{E}_i^{\downarrow}
                   + W_i^{\uparrow}(T) \mathcal{E}_i^{\uparrow}\;,
\end{equation}
with the two rates $W_i^{\downarrow}(T)=(1-T)\lambda_{\text{B}}$ and $W_i^{\uparrow}(T)=T\lambda_{\text{B}}$ ($\lambda_{\text{B}}$ is the coupling strength of the environment, $T$ its temperature) and $\mathcal{E}_{i}^{\downarrow}$, $\mathcal{E}_{i}^{\uparrow}$ are transition operators.

With this definition the perturbed super operator (\ref{eq:7}) can be rewritten as
\begin{equation}
  \label{eq:15}
  \mathcal{L'}(\Delta T) 
  = \mathcal{L}_{1}(T)+\mathcal{L}_{2}(T)
  + \frac{\Delta T \lambda_{\text{B}}}{2} \mathcal{E}\;,
\end{equation}
with $\mathcal{E}=-\mathcal{E}_{1}^{\downarrow}+\mathcal{E}_{1}^{\uparrow}+\mathcal{E}_{2}^{\downarrow}-\mathcal{E}_{2}^{\uparrow}$.
The first two terms are just bath systems at temperature $T$ like in the unperturbed case, we therefore neglect these terms in the following.
Like in standard perturbation theory in Hilbert space we can calculate the state correction $\Delta\hat{\rho}$ in first order perturbation theory with respect to the stationary state of the unperturbed system $\hat{\rho}_0$ (keeping in mind that the eigensystem is not orthogonal) as
\begin{equation}
  \label{eq:27}
  \hat{\rho}_{\text{stat}} = \hat{\rho}_0 + \Delta\hat{\rho} = 
    \hat{\rho}_0 - \frac{\Delta T\lambda_{\text{B}}}{2}
    \sum_{j=1}^{n^{2N}-1}
    \frac{\braL{\hat{\rho}^j} \mathcal{E} \ketL{\hat{\rho}_0}}{l_j}\,
    \ketL{\hat{\rho}_j}\;.
\end{equation}
For a more detailed derivation of this formula see \cite{Michel2004}.

% ---------------------------------------------
%
% Fifth Chapter:
%
\section{Investigation of the Stationary State}
\label{sec:level5}

Since the stationary density operator of the system is now given by $\hat{\rho}=\hat{\rho}_0+\Delta\hat{\rho}$ and since we know that $\hat{\rho}_0$ does not give rise to any local temperature difference and current, the expectation values of the operators defined above are determined only by $\Delta\hat{\rho}$ from (\ref{eq:27}).
We thus find for the local internal temperature gradient 
\begin{align}
  \label{eq:23}
  &\delta T^{(\mu,\mu+1)} 
   = \trtxt{\Delta\hat{H}^{(\mu,\mu+1)}_{\text{loc}} \Delta\hat{\rho}}\notag\\
  &\quad= -\frac{\Delta T\lambda_{\text{B}}}{2} \sum_{j=1}^{n^{2N}-1}
     \frac{\braL{\hat{\rho}^j} \mathcal{E} \ketL{\hat{\rho}_0}}{l_j}\,
     \trtxt{\Delta\hat{H}^{(\mu,\mu+1)}_{\text{loc}}\hat{\rho}_j}
\end{align}
and for the local current within the system
\begin{align}
  \label{eq:33}
  &J^{(\mu,\mu+1)} 
   = \tr{\hat{J}^{(\mu,\mu+1)} \Delta\hat{\rho}}\notag\\
  &\quad= -\frac{\Delta T\lambda_{\text{B}}}{2} \sum_{j=1}^{n^{2N}-1}
     \frac{\braL{\hat{\rho}^j} \mathcal{E} \ketL{\hat{\rho}_0}}{l_j}\,
     \trtxt{\hat{J}^{(\mu,\mu+1)}\hat{\rho}_j}\;.
\end{align}
The current as well as the local temperature gradient depend linearly on the global temperature difference of the bath systems.
Under stationary conditions the current must be independent of $\mu$, $J^{(\mu,\mu+1)}=J$, so that (\ref{eq:33}) can be rewritten as
\begin{equation}
  \label{eq:34}
  J = - \kappa' \Delta T\;.
\end{equation}
Eigenstates and eigenvalues entering here the \emph{global} conductivity $\kappa'$ depend only on the mean temperature of the unperturbed system, not on $\Delta T$.
Based on this $\kappa'$ as a global property of the system, including its contact properties to the environments, let us call (\ref{eq:34}) ``external Fourier's Law''.  

Furthermore, combining (\ref{eq:23}) and (\ref{eq:33}), we can define a \emph{local} conductivity within the system 
\begin{equation}
  \label{eq:21}
  \kappa^{(\mu,\mu+1)} 
  = -\frac{J^{(\mu,\mu+1)}}{\delta T^{(\mu,\mu+1)}}
  = -\frac{J}{\delta T^{(\mu,\mu+1)}}
\end{equation}
implying also $\kappa^{(\mu,\mu+1)}$ to be independent of the external gradient $\Delta T$.

% ---------------------------------------------
%
% Sixth Chapter:
%
\section{Numerical Investigation of the System}
\label{sec:level6}

\begin{figure}
  \centering  
  \psfrag{kappa}{$\kappa^{(2,3)}$}
  \psfrag{DeltaT}{\raisebox{-4pt}{$\Delta T$}}
  \psfrag{das ist der erste}{\hspace{-2mm}\tiny eq. (\ref{eq:8})}
  \psfrag{das ist der zweite}{\hspace{-2mm}\tiny eq.\ (\ref{eq:21})}
  \includegraphics[width=5cm]{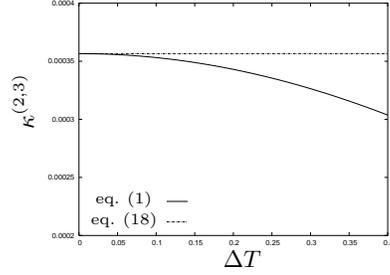}
  \caption{Local conductivity $\kappa^{(2,3)}$ in a Heisenberg spin chain of 4 spins as a function of the external perturbation $\Delta T$; the solid line refers to the solution of the full LvN equation, the dashed line shows (\ref{eq:21}).}
  \label{fig:1}
\end{figure}

We compare these result with the complete numerical solution of the LvN equation (\ref{eq:8}), here for a Heisenberg spin chain with four spins.
In Fig.~\ref{fig:1} we show the local conductivity of the two central spin systems $\kappa^{(2,3)}$ as a function of the external gradient $\Delta T$.
Indeed, we find numerically that $\kappa^{(2,3)}$ according to (\ref{eq:21}) does not depend on $\Delta T$ (dashed line).
The exact numerical solution of the LvN equation shows a weak $\Delta T$ dependence (solid line).
As expected, linear transport and our perturbational theory apply for not too large external gradients $\Delta T$ only.
Note that the full range of temperatures $0\leq T<\infty$ has been mapped here onto the interval $[0,0.5]$ ($\Delta T = 0.3$ is thus already a large gradient). 

We address now some remarkable features of very small heat conducting model systems, based on the numerical solution of the complete LvN equation (compare \cite{Michel2003}).

\begin{figure}\centering 
  \psfrag{T}{\hspace{-0.5cm}\raisebox{3pt}{$T[\Delta E]$}}
  \psfrag{mu}{\hspace{-1cm}\raisebox{-5pt}{system number $\mu$}}
  \psfrag{T1}{}
  \psfrag{T2}{}
  \includegraphics[width=5cm]{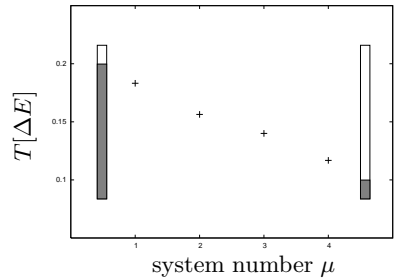}
  \caption{Temperture profile of a four spin Heisenberg chain, as calculated by the full LvN equation. Within the chain we find a constant gradient and therefore normal heat conduction.}
  \label{fig:2}
\end{figure}

\begin{figure}\centering 
  \psfrag{T}{\hspace{-0.5cm}\raisebox{3pt}{$T[\Delta E]$}}
  \psfrag{mu}{\hspace{-1cm}\raisebox{-5pt}{system number $\mu$}}
  \psfrag{T1}{}
  \psfrag{T2}{}
  \includegraphics[width=5cm]{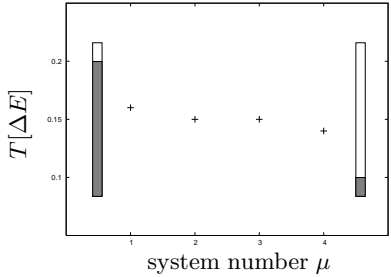}
  \caption{Temperture profile of a four spin XY model, as calculated by the full LvN equation. Within the chain we find a vanishing gradient and therefore ballistic transport.}
  \label{fig:3}
\end{figure}

``Normal'' heat conduction (Fourier's Law) is associated with a constant but non-zero local temperature gradient and thus a finite conductivity $\kappa^{(\mu,\mu+1)}=\kappa$ independent of site $\mu$ everywhere in the system (see \cite{Mahan1981,Gemmer2004}).
In Fig.~\ref{fig:2} we show the temperature profile of the Heisenberg spin chain for different coupling types within the chain. 
The data are obtained from the full solution of (\ref{eq:8}).
The majority of coupling types, the Heisenberg coupling and the random next neighbor interaction, indeed show the normal behavior in the weak coupling limit as demonstrated in Fig.~\ref{fig:2} (\cite{Michel2003,Gemmer2004}).
But this ``normal'' transport type does not always show up:
A spin-spin interaction consisting of an energy transfer coupling only (XY model) leads to a vanishing temperature gradient off the contact regions (see Fig.~\ref{fig:3}).

The vanishing gradient in Fig.~\ref{fig:3} implies a divergent conductivity within the chain $\kappa^{(\mu,\mu+1)}$.
Nevertheless the current remains finite because of the resistance at the contacts, therefore the global conductivity $\kappa$ defined in (\ref{eq:34}) also remains finite for this special coupling type.
Therefore we could state that the ``external Fourier's Law'' is valid even if Fourier's Law proper does not apply.

\begin{figure}
  \centering
  \psfrag{Strom}{\hspace{-1cm}\raisebox{3pt}{$\scriptscriptstyle J^{(2,3)}[10^{-7}\Delta E]$}}
  \psfrag{DeltaT}{\hspace{-.5cm}\raisebox{-6pt}{$\Delta T^{(2,3)}[\Delta E]$}}
  \psfrag{Temp1}{\tiny\hspace{-1.8cm}\raisebox{-1.2pt}{$T_{\text{mean}}=0.00015$}}
  \psfrag{Temp2}{\tiny\hspace{-1.8cm}\raisebox{-1.2pt}{$T_{\text{mean}}=0.2015$}}
  \psfrag{Temp3}{\tiny\hspace{-1.8cm}\raisebox{-1.2pt}{$T_{\text{mean}}=0.4515$}}
  \includegraphics[height=5cm,angle=-90]{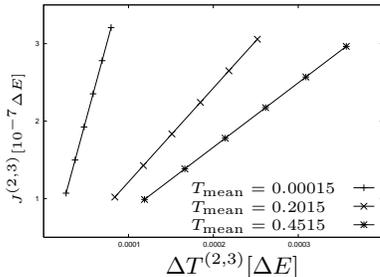}
  \caption{Fourier's Law in a four spin Heisenberg chain. For different mean temperatures the current is linear in the local gradient. Conductivity depends on the mean temperature.}
  \label{fig:4}
\end{figure}

\begin{figure}
  \centering
  \psfrag{kappa}{$\kappa$}
  \psfrag{TMittel}{\hspace{-.5cm}\raisebox{-6pt}{$T_{\text{mean}}[\Delta E]$}}
  \includegraphics[height=5cm,angle=-90]{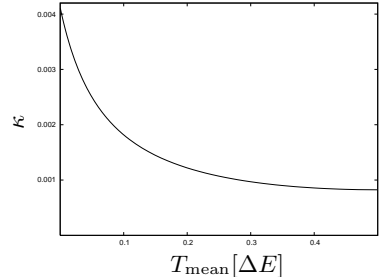}
  \caption{Mean temperature dependence of local conductivity.}
  \label{fig:5}
\end{figure}

For the current, in case of a normal transport behavior, in dependence of the internal gradient we find a linear behavior, like proposed by Fourier's law shown for three different mean temperatures of the bath systems in Fig.~\ref{fig:4}.
Obviously, the \emph{local} conductivity depends on the mean temperature of the bath systems.
Furthermore, we show the dependence of the conductivity on mean temperature (see Fig.~\ref{fig:5}).

By comparing these solutions of the full LvN equation with the results of the perturbation theory we find a very good accordance for not to large gradients, as expected.
The perturbation theory reflects the vanishing gradient in energy exchange coupling as well as the normal linear gradient for Heisenberg chains. 
Also the conductivities are the same as found in the full solution of LvN equation.

% ----------------------------------------------------------------------------
%
% 7. Chapter:
%
\section{Conclusion}
\label{sec:level7}

We have considered heat conduction in small quantum systems built up from identical subsystems weakly coupled by some next neighbor interaction.
By a perturbation theory similar to that introduced by Kubo but extended to the full Liouville space of the system, we have been able to derive a quantitative equation for the temperature gradient and heat currents within such systems. 
These equations  depend only on properties of the unperturbed system and linearly on the strength of the perturbation $\Delta T$.

The most remarkable point of the equation for the heat current and the temperature profile is the fact that the global temperature difference of the external bath systems shows up only as a parameter.
This is not only a numerical advantage, since a diagonalization for different global gradients is no longer necessary, but also an interesting physical fact:
The heat conductivity is independent of the external gradient $\Delta T$, and the ``external Fourier's Law'' is always fulfilled, even if the internal gradient of the system is not constant, as long as the perturbation theory applies. 

Our approach does not have the problem of introducing a potential term into the Hamiltonian of the system, like in standard Kubo formulas for heat conduction.
The bath systems, modeled by a Linblad formalism, directly define the perturbation in Liouville space.
Like in standard perturbation theory in Hilbert space, the first order correction to the stationary state of the system is expressed in terms of transition matrix elements of the perturbation operator and the eigenstates and eigenvalues of the unperturbed system.
Only the non-orthogonality of the eigensystem of the unperturbed system needs a more careful treatment, formally the equations are very similar. 

Investigations on further aspects of the derived formalism are in progress.
We hope to clarify the question of the different transport behavior (non-vanishing and vanishing local gradients) under different coupling types.

% -----------------------------------------------------------------------------
%
% The End of the Main Text
%
% -----------------------------------------------------------------------------

We thank M.\ Hartmann, M.\ Henrich, Ch.\ Kostoglou, H.\ Michel, H.\ Schmidt, M.\ Stollsteimer and F.\ Tonner for fruitful discussions. Financial support by the Deutsche Forschungsgesellschaft is gratefully acknowledged.

%\bibliographystyle{elsart-num}
%\bibliography{qthermo}

\end{document}